\documentclass[twocolumn]{aastex63}
\usepackage{longtable}
\usepackage{graphicx}
\usepackage{float}
\usepackage{amsmath,amssymb}
\usepackage{xcolor}
\usepackage{hhline}
\usepackage{multirow}
 \usepackage{textcomp, gensymb} 
\usepackage{soul} 
\usepackage{colortbl}

\usepackage{colortbl}

\usepackage{natbib}
\received{}
\revised{}
\accepted{}
\submitjournal{ApJ}
\shorttitle{Jets in  FXTs}
\shortauthors{Dina Ibrahimzade}
\graphicspath{{./}{figures/}}
\begin{document}
\title{Constraints on Relativistic Jets from the Fast X-ray Transient 210423 using Prompt Radio Follow-up Observations}
\correspondingauthor{Dina Ibrahimzade}
\email{dinaevazade@berkeley.edu }
\author[0009-0008-0782-5028]{Dina Ibrahimzade}
\affiliation{Department of Astronomy, University of California, Berkeley, CA 94720-3411, USA}
\author[0000-0003-4768-7586]{R. Margutti}
\affiliation{Department of Astronomy, University of California, Berkeley, CA 94720-3411, USA}
\affiliation{Department of Physics, University of California, 366 Physics North MC 7300,
Berkeley, CA 94720, USA}
\author[0000-0002-7735-5796]{J.~S. Bright} 
\affiliation{Astrophysics, Department of Physics, University of Oxford, Denys Wilkinson Building, Keble Road, Oxford OX1 3RH, UK}
\affiliation{Department of Astronomy, University of California, Berkeley, CA 94720-3411, USA}
\author[0000-0003-0526-2248]{P. Blanchard} 
\affiliation{Center for Interdisciplinary Exploration and Research in Astrophysics (CIERA), Northwestern University, Evanston, IL 60202, USA}
\affiliation{Department of Physics and Astronomy, Northwestern University, Evanston, IL 60208, USA}
\author[0000-0001-8340-3486]{K.  Paterson} 
\affiliation{Max-Planck-Institut f\"ur Astronomie, K\"onigstuhl 17, 69117 Heidelberg, Germany}
\author[0000-0001-5683-5339]{D. Lin} 
\affiliation{Department of Physics, Northeastern University, Boston, MA 02115-5000, USA}
\author[0000-0001-8023-4912]{H. Sears} 
\affiliation{Center for Interdisciplinary Exploration and Research in Astrophysics (CIERA), Northwestern University, Evanston, IL 60202, USA}
\affiliation{Department of Physics and Astronomy, Northwestern University, Evanston, IL 60208, USA}
\author[0000-0002-5283-933X]{A. Polzin}
\affiliation{Department of Astronomy and Astrophysics, The University of Chicago, Chicago, IL 60637, USA} 
\author[0000-0002-8977-1498]{I. Andreoni}
\affiliation{Joint Space-Science Institute, University of Maryland, College Park, MD 20742, USA}
\affiliation{Department of Astronomy, University of Maryland, College Park, MD 20742, USA}
\affiliation{Astrophysics Science Division, NASA Goddard Space Flight Center, Mail Code 661, Greenbelt, MD 20771, USA}
\author[0000-0001-9915-8147]{G.~Schroeder}
\affiliation{Center for Interdisciplinary Exploration and Research in Astrophysics (CIERA), Northwestern University, Evanston, IL 60202, USA}
\affiliation{Department of Physics and Astronomy, Northwestern University, Evanston, IL 60208, USA}
\author[0000-0002-8297-2473]{K. D. Alexander} 
\affiliation{Department of Astronomy/Steward Observatory, 933 North Cherry Avenue, Rm. N204, 
Tucson, AZ 85721-0065, USA}
\author[0000-0002-9392-9681]{E. Berger} 
\affiliation{Center for Astrophysics | Harvard \& Smithsonian, 60 Garden Street, Cambridge, MA 02138-1516, USA}
\author[0000-0001-5126-6237]{D.~L. Coppejans} 
\affiliation{Department of Physics, University of Warwick, Gibbet Hill Road, CV4 7AL Coventry, United Kingdom}
\author[0000-0003-2349-101X]{A. Hajela} 
\affiliation{DARK, Niels Bohr Institute, University of Copenhagen, Jagtvej 128, 2200 Copenhagen, Denmark}
\author[0000-0003-4307-8521]{J. Irwin} 
\affiliation{Department of Physics \& Astronomy, University of Alabama, Tuscaloosa, AL 35487-0324, USA}
\author[0000-0003-1792-2338]{T. Laskar} 
\affiliation{Department of Physics \& Astronomy, University of Utah, Salt Lake City, UT 84112, USA}
\author[0000-0002-4670-7509]{B.~D. Metzger} 
\affiliation{Department of Physics, Columbia University, New York, NY 10027, USA}
\affiliation{Center for Computational Astrophysics, Flatiron Institute, 162 5th Avenue, New York, NY 10010, USA}
\author[0000-0002-9267-6213]{J.~C. Rastinejad}  
\affiliation{Center for Interdisciplinary Exploration and Research in Astrophysics (CIERA), Northwestern University, Evanston, IL 60202, USA} 
\affiliation{Department of Physics and Astronomy, Northwestern University, Evanston, IL 60208, USA}
\author[0000-0003-2705-4941]{L. Rhodes}
\affiliation{Astrophysics, Department of Physics, University of Oxford, Denys Wilkinson Building, Keble Road, Oxford OX1 3RH, UK}

\begin{abstract}
Fast X-ray Transients (FXTs) are a new observational class of phenomena with no clear physical origin. This is at least partially a consequence of limited multi-wavelength follow up of this class of transients in real time. Here we present deep optical ($g-$ and $i-$ band) photometry with Keck, and prompt radio observations with the VLA of FXT\,210423 obtained at $\delta t \approx 14-36$\,days since the X-ray trigger. We use these multi-band observations, combined with publicly available data sets, to constrain the presence and physical properties of on-axis and off-axis relativistic jets such as those that can be launched by neutron-star mergers and tidal disruption events, which are among the proposed theoretical scenarios of FXTs. Considering a wide range of possible redshifts $z\le3.5$, circumstellar medium (CSM) density $n=10^{-6}-10^{-1}\,\rm{cm^{-3}}$, isotropic-equivalent jet kinetic energy $E_{k,iso}=10^{48}-10^{55}\,\rm{erg}$, we find that we can rule out wide jets with opening angle $\theta_{j}=15\degree$ viewed within $10\degree$ off-axis. For more collimated jets ($\theta_{j}=3\degree$) we can only rule out on-axis ($\theta_{obs}=0\degree$) orientations. This study highlights the constraining power of prompt multi-wavelength observations of FXTs discovered in real time by current (e.g., Einstein Probe) and future facilities.
\end{abstract}


\hfill
\section{Introduction}
\label{sec:intro} 
Fast X-Ray Transients (FXTs) are flashes of X-ray emission with time scales between hundreds of seconds to hours and a remarkably large range of intrinsic luminosities spanning several decades (see e.g., \citealt{polzin2023luminosity}, their figure 5). Identified in the past decade (\citealt{Soderberg08,Jonker13,Glennie15,Irwin16,DeLuca16,bauer2017new,Novara20SBO}),  FXTs represent an emerging new and heterogeneous observational class of phenomena not well explained by any single model. 

Several scenarios have been put forth to explain the physical origin of these events, including both Galactic and extra-Galactic origins. Among these is the possibility that FXTs represent the emission originating from the shock break out (SBO) from a core-collapse supernova (CC-SN), as a flash of X-ray to UV radiation is expected when the shock wave crosses the surface of the star (e.g., \citealt{Waxman17}). A possible SBO origin of some FXTs has been suggested by \cite{Soderberg08,Novara20SBO,Alp2020,Eappachen24}, although we note that the X-Ray Transient XRT\,080901\footnote{The nomenclature ``FXT'' did not exist at the time of discovery of this transient, hence the ``XRT'' (or ``XRO'', i.e., X-Ray Outburst) name.}  is the only case with a spectroscopically confirmed optical supernova (i.e., SN\,2008D; \citealt{Soderberg08,Mazzali08,Modjaz09}).

Other proposed interpretations are related to low-luminosity long gamma-ray bursts or short gamma-ray bursts (LGRBs or SGRBs, respectively) viewed off axis  (e.g., \citealt{Jonker13,Glennie15,bauer2017new,sarin2021cdf, EP240315_levan2024fast}). The possible connection with manifestations of SGRB progenitors implies that some FXTs could be produced by a rapidly-spinning magnetar remnant that resulted from the merger of two neutron stars (NSs) (e.g., \citealt{xue2019magnetar,sun2019unified,ai_binary_post_merger_2021,lin2022discovery,eappachen2023fast,Eappachen24}). Mergers between a white dwarf (WD) and an NS represent another interpretation. The properties of the expected transients from these events are not well constrained due to the broad range of physical processes involved \citep{Fernandez19}. 

Tidal disruption events (TDEs), where the X-rays are emitted as a consequence of the accretion of gas that results from the shredding of a star as it passes too close to a black hole (BH), represent another potential origin. Specifically, given the short time-scales of evolution of FXTs, the tidal disruption of WDs on Intermediate Mass BHs (WD-IMBH TDEs, e.g., \citealt{MacLeod16, Maguire2020}) has been proposed as a possible scenario (e.g., \citealt{Jonker13,Glennie15,Irwin16,bauer2017new,Peng19}). Directly relevant to our study, GRBs and TDEs can launch relativistic jets \citep{Andreoni_jetted_TDE_rare, Anreoni_22_jetted_TDE_lum, Rhodes_TDE, Pasham_TDE, Somalwar_TDE, Bloom_TDE, Burrows_TDE, Levan_TDE, Zauderer_TDE, Cenko_TDE, Brown_TDE, Yuan_TDE}.

Other scenarios include emission from X-ray binaries, and Galactic phenomena such as late-type stellar flares (e.g., \citealt{Glennie15,DeLuca2020}) as it was suggested for some XMM-\emph{Newton} FXTs (\citealt{Alp2020}).
Finally, magnetar outbursts and even exotic scenarios involving the collisions of minor bodies (such as asteroids) with NSs  have been considered (e.g., \citealt{Jonker13}). Understanding the origins and progenitors of FXTs is paramount and may also have implications for current gravitational wave (GW) searches (i.e., if indeed associated with binary NS mergers, FXTs could constitute electromagnetic counterparts of GW sources). 

While approximately 40 FXTs have been documented up to the end of 2023,
(see e.g., \citealt{quirola2022extragalactic,QuirolaVasquez23,Eappachen24} for recent population studies, and references therein), the vast majority have been discovered through archival data searches (mostly using Chandra or XMM-\emph{Newton} data) and therefore lack prompt multi-wavelength follow up, making their characterization and determination of their physical origin challenging. With the recent launch of the Einstein Probe (EP) in early 2024 \citep{yuan2022einstein}, the  detection of FXTs in real time has now become routine. This opens up the possibility of prompt follow up, which is likely to shed new light on the origins of these events.

A notable example in this regard is the recently-discovered EP\,240315 in the 0.5-4 keV band by \cite{EP240315_detection}. Follow up of the event led to the first detection of both an optical and radio counterpart to an FXT and a redshift measurement of $z=4.859$ \citep{EP240315_radioFXTdiscovery, EP240315_opticalgcn, EP240315_redshiftgcn, EP240315_MeerKAT,EP240315_eMERLIN, EP240315_ATCA}. The observed properties of EP\,240315 are consistent with an interpretation as an LGRB and illustrate the possibility that a large fraction of FXTs with lower X-ray luminosities may be interpreted as low-luminosity LGRBs \citep{EP240315_radioFXTdiscovery, EP240315_levan2024fast, EP240315_liu2024soft}.  \cite{EP240315_radioFXTdiscovery} also recognize that with the current observations, a jetted TDE interpretation cannot be ruled out.

The EP detection of similar events at cosmological distances and in the local Universe will likely continue to shed insight on this potential categorization. 

In this paper, we focus on FXT\,210423, which was discovered in  real time by \cite{lin2021chandra} in Chandra data. Largely based on the observed properties of the X-ray signal, \cite{ai_binary_post_merger_2021} and \cite{lin2021chandra} propose that FXT\,210423 is the manifestation of a rapidly-spinning magnetar remnant resulting from a merger of two NSs, like SGRBs. \cite{eappachen2023fast} present a detailed analysis of the X-ray light curve and spectrum of FXT\,210423 along with deep optical imaging of the event, and identify three potential host galaxies. While acknowledging multiple possible origins of the event including a WD-IMBH TDE, SBO, WD-NS merger, or NS-NS merger (including the possibility of a remnant magnetar), \cite{eappachen2023fast} favour a NS merger scenario. In this work we present deep optical and radio follow up observations of FXT\,210423. We use these observations in combination with simulations to place constraints on the parameter space of relativistic jets in effort to narrow down the the possible origins of the transient and provide future recommendation for prompt multi-wavelength follow up of FXTs. 

This paper is organized as follows: in  \S\ref{sec:data} we discuss the extraction of the X-ray, radio, and optical data used for our analysis, and we discuss the possible host galaxy association of the event using deep optical imaging. In \S\ref{sec:simulations} we derive constraints on the presence of relativistic jets like those launched by SGRBs, LGRBs or TDEs using a large sample of jet simulations. We discuss our findings in \S\ref{sec:disc},
and we comment on the implications of these results for future follow-up campaigns of FXTs discovered in real-time, for example with the EP. We adopt the following cosmology: $H_{0}$= 69.6 km s$^{-1}$ Mpc$^{-1}$, $\Omega_{M}=0.286$, and $\Omega_{\Lambda}=0.714$.

\begin{figure}[!ht]
\hskip -0.5 cm
\includegraphics[width=.5\textwidth]{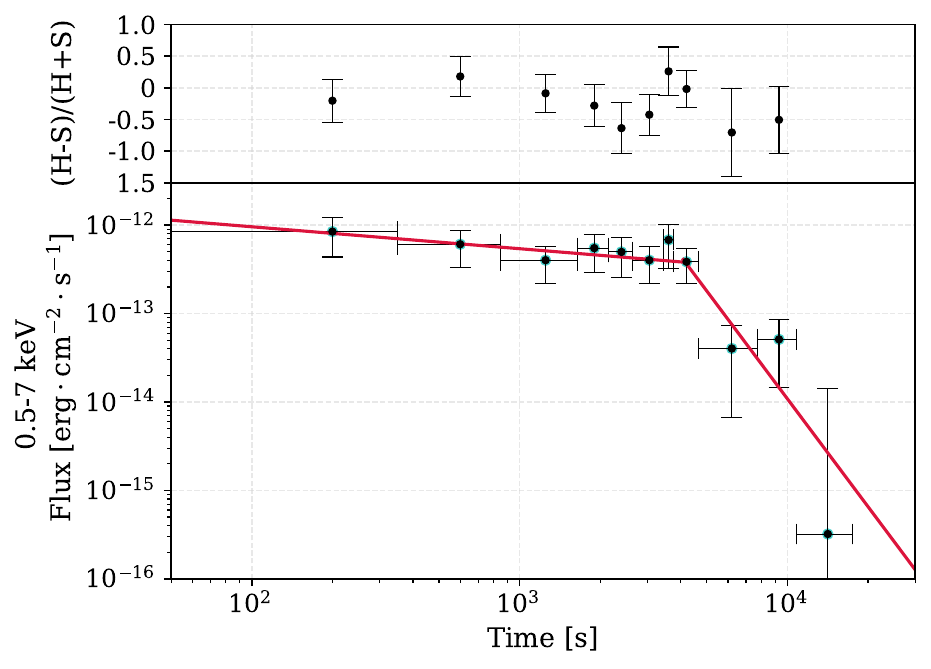}
\caption{Unabsorbed 0.5-7 keV flux (\emph{main panel}) and hardness ratio (\emph{upper panel}) evolution of FXT\,210423.  
The soft band (S) is defined between 0.5-2 keV, and the hard band (H) between  2-7 keV. No spectral evolution is apparent. The X-ray light-curve can be fit with a broken power law (red solid line)  with $F_{X} \propto t^{-0.2 \pm 0.2}$ at $\delta t<4.1\,\rm{ks}$ steepening to $F_{X} \propto t^{-4.1 \pm 1.4}$
at later times.}
\label{figure:LC}
\end{figure}

\section{Observations and Data analysis} \label{sec:data}
\subsection{Chandra X-ray Observations}
FXT\,210423 was detected on April 23 2021 (MJD 59327) at 22:22:35.817 UTC until April 24 2021 at 02:15:05.817 UTC, during a Chandra X-ray Observatory (CXO) calibration observation of Abell 1795 (ObsID:24604) \citep{lin2021chandra}. The source is located on CCD ID3 of the ACIS-I instrument. We reprocessed and reduced the data using the \texttt{CIAO} v4.15.1 software package and standard filtering criteria with the goal to obtain the X-ray flux light-curve of FXT\,210423. While the CXO data analysis has already been presented in  \cite{eappachen2023fast}, a table of corresponding time and flux values of the light curve has not previously been published. 

\begin{deluxetable}{cccc}
\tablehead{
\colhead{MJD} \vspace{-0.2cm} & \colhead{Time since $T_{0}$}& \colhead{Flux  } &  \\ (d) & \colhead{(s)} &  \colhead{$(\rm{erg}\,  \rm{cm}^{-2} \rm{s}^{-1})$}}
\tablecaption{Chandra  unabsorbed 0.5-7 keV flux light curve of FXT\,210423. Light curve shown in Fig. \ref{figure:LC}. \label{table:Xray}}
\startdata
59327.9324 & 200 &  $8.44^{+3.77}_{-4.08}\times 10^{-13}$\\
59327.9370 & 600 &  $6.08^{+2.53}_{-2.76}\times 10^{-13}$\\
59327.9445 & 1250 &  $4.01^{+1.66}_{-1.80}\times 10^{-13}$\\
59327.9520 & 1900 &  $5.49^{+2.39}_{-2.59}\times 10^{-13}$\\
59327.9578 & 2400 &  $4.97^{+2.25}_{-2.44}\times 10^{-13}$\\
59327.9653 & 3050 &  $4.01^{+1.66}_{-1.80}\times 10^{-13}$\\
59327.9717 & 3600 &  $6.80^{+3.29}_{-3.54}\times 10^{-13}$\\
59327.9787 & 4200 &  $3.85^{+1.54}_{-1.68}\times 10^{-13}$\\
59328.0018 & 6200 &  $4.02^{+3.21}_{-3.36}\times 10^{-14}$\\
59328.0377 & 9300 &  $5.10^{+3.46}_{-3.65}\times 10^{-14}$\\
59328.0938 & 14150 &  $3.22^{+1.38}_{-1.38}\times 10^{-16}$\\
\enddata
\end{deluxetable}

\begin{figure}
\includegraphics[width=.5\textwidth]{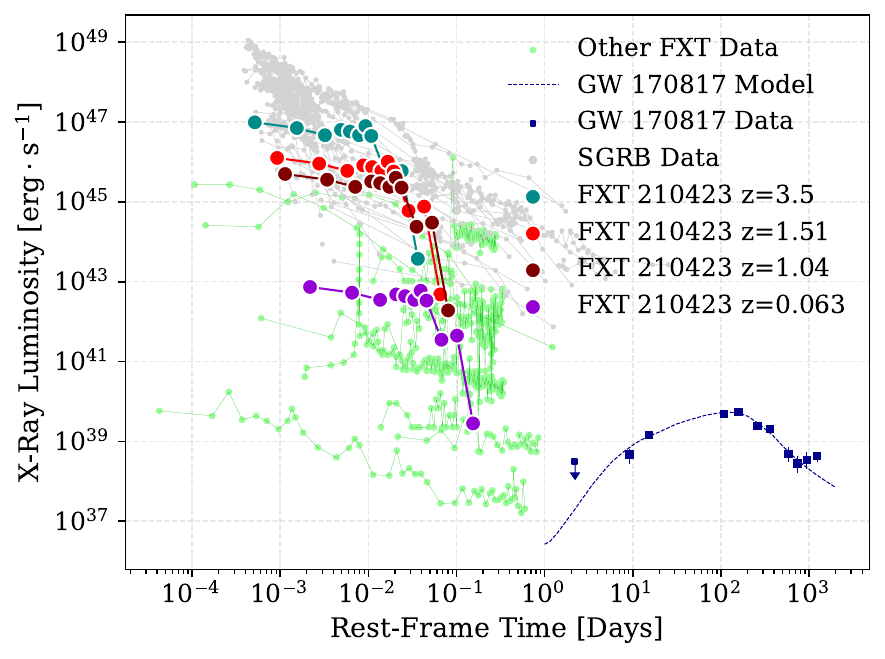}
\caption{Comparison of the 0.3-10 keV (observer frame) X-ray luminosity vs. rest frame time of FXT\,210423 assuming different redshift values,  with the X-ray emission from the NS merger event GW\,170817 from \cite{hajela2022evidence}, a collection of  SGRB X-ray afterglows from \cite{rouco2022jet}, and a collection of FXT X-ray light curves from \cite{polzin2023luminosity}. The X-ray luminosity of FXT\,210423  at $z\ge1$ (\S\ref{sec:simulations}) is consistent with that of SGRB afterglows and the more luminous FXTs.}
\label{figure:X_Ray_Comparison}
\end{figure}

We used the \texttt{CIAO} tool \texttt{wavdetect} and we detected an X-ray source with significance $\sigma = 10.89$ at sky coordinates  $\rm{RA}=13^{h}48^{m}56.46^{s}\pm 0.04^{s}$ DEC: $\delta=26\degree 39^{'}45.13{''}\pm 0.54{''}$ consistent with those reported in \cite{eappachen2023fast}. Consistent with previous findings, the X-ray emission from FXT\,210423 is detected by the CXO for a total of 13.95 ks.   
Based on the output from \texttt{wavdetect}, we used a source extraction region centered at the coordinates above and with radius of $7.5^{''}$, while we estimated the background contribution using a $60^{''}$ radius circular source-free region. We note that the  off-axis source location in the telescope field of view causes a significantly larger than average  PSF.  The source region contains a total of 140 events  in the energy range 0.5-7 keV, corresponding to 108.2 background-subtracted events. 
We extracted the net count-rate light-curves of the source in the energy ranges 0.5-2 keV (soft band) and 2-7 keV (hard band) with \texttt{dmextract}. As we show  in Fig. \ref{figure:LC}, upper panel, we find no statistical evidence for evolution of the hardness ratio with time, consistent with \cite{eappachen2023fast}. Given the  lack of evidence for spectral evolution of the source and the limited photon statistics, we proceeded with the extraction of a single spectrum with \texttt{specextract}. 
We modeled the spectrum as an absorbed power-law model (i.e., \texttt{tbabs*ztbabs*pow} within \texttt{Xspec}). We set the Galactic neutral hydrogen column density in the direction of the source to $N_\mathrm{H,MW} = 1.01 \times 10^{20} \rm{cm^{-2}}$ \citep{HI4PI}. We found no evidence for intrinsic absorption, and we thus assumed $N_\mathrm{H,int} = 0$ $\rm{cm^{-2}}$.  For this paper we explore a range of potential source redshifts $z=$ 0.063, 1.0, 1.5, 3.5 (section \ref{sec:simulations}), but we note that the best-fitting X-ray spectral parameters do not significantly depend on $z$. 
We employed Cash-statistics for our fits. The best-fitting power-law photon index is $\Gamma=1.90^{+0.28}_{-0.27}$, where $\Gamma=\beta +1$ and the specific flux is $F_{\nu}\propto \nu^{-\beta}$. We used this model to flux calibrate the 0.5-7 keV net count-rate light-curve in the 0.5-7 keV energy band. We provide the unabsorbed 0.5-7 keV flux light-curve of FXT\,210423 in Table \ref{table:Xray} and we show the X-ray flux evolution of  FXT\,210423 in Fig. \ref{figure:LC}.  We show the comparison of the X-ray luminosity at four assumed redshift values for FXT \,210423 with a collection of FXT X-ray light curves, a collection of SGRB X-ray afterglows, and the X-ray emission from the NS merger event GW\,170817 in Fig.\ref{figure:X_Ray_Comparison}. The results from our X-ray analysis are consistent with those from \cite{eappachen2023fast}. 

\subsection{Radio Observations} 
\begin{figure}[!ht]

\includegraphics[width=0.5
\textwidth]{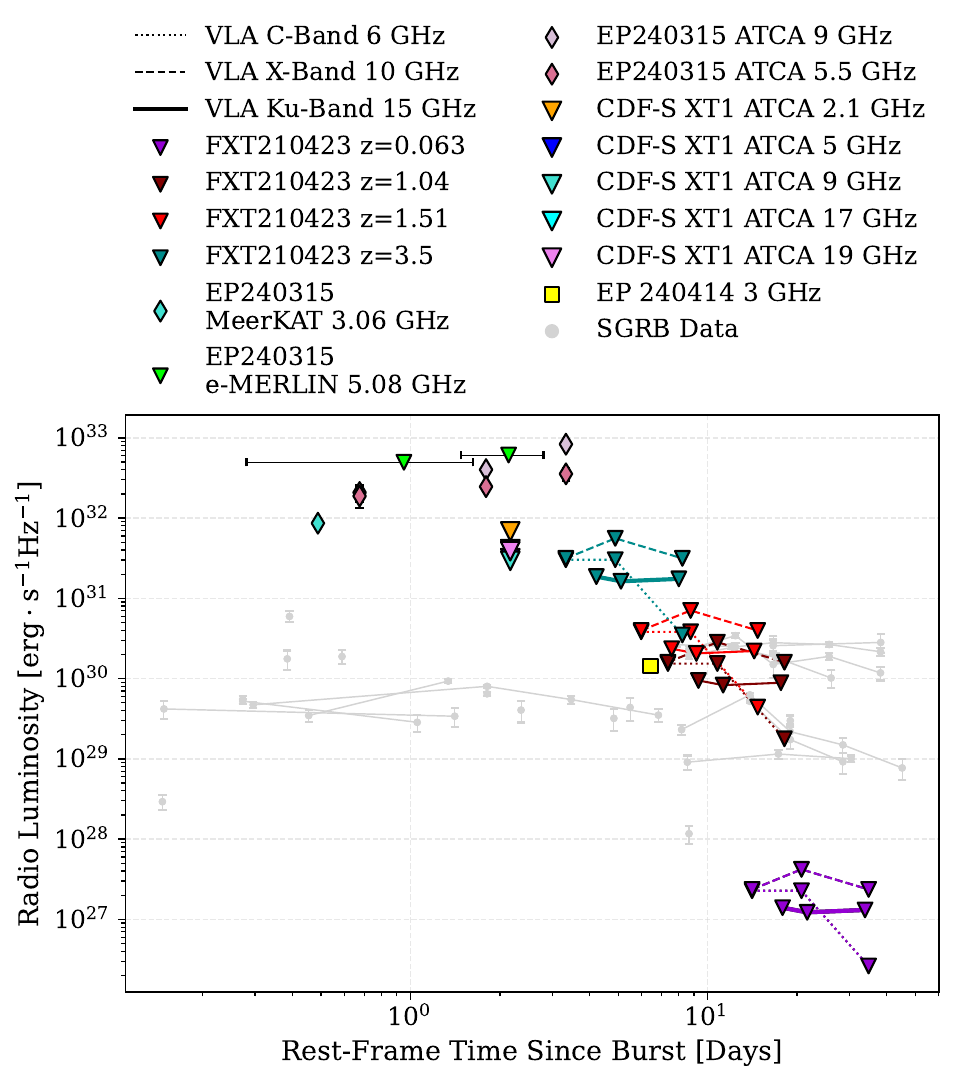}
\caption{Radio observations of FXT\,210423 for four assumed redshift values in the context of radio afterglows of  SGRBs at rest-frame frequency $6-15$ GHz and other FXTs with prompt radio observations. Triangles represent $3 \sigma$ upper limits while all other markers represent detections.  Radio observations of SGRBs from \cite{GRB051221A, GRB050724, GRB141212A, GRB130603B,  GRB200522A, GRB211106A, Z_sources_1, GRB170728, GRB051221A_z, Z_sources_2, Z_sources_3}. EP240315 data from \cite{EP240315_radioFXTdiscovery}. EP240414 data from \citep{EP240414_Radio, EP240414_Optical}.  CDF-S FXT1 data from \cite{bauer2017new}. SRGt J123822.3-253206, a source consistent with the features of an FXT  with radio counterpart \citep{ATel13415, ATel13416, ATel13485}, is not shown in this plot due to the lack of a known redshift.  Radio luminosity of FXT\,210423 at $z\ge 1$ is consistent with that of the population of SGRB radio afterglows. 
}
\label{figure:RadioFXT}
\end{figure}

We initiated Karl G. Jansky Very Large Array (VLA) radio observations of XRT\,210423 under the DDT program VLA/21A-422 (PI J. Bright) beginning on May 08 2021. Observations were taken with the VLA in its most compact (D) configuration, and at C-, X-, and Ku-band using the WIDAR backend for optimal continuum sensitivity. Data were calibrated using the Common Astronomy Software Applications (CASA; \citealt{mcmulling2007}) version 6.4.1.12 pipeline version 2022.2.0.64, with  3C286 used to calibrate the bandpass response of the VLA and the absolute flux scale, while J1407$+$2827 was used to calibrate the time dependent gains at all frequencies. The pipeline output was then imaged manually using the CASA task \textsc{tclean} with a briggs robust parameter of 0.5 \citep{briggs1995}. Upon imaging the calibrated data we identify significant residual artefacts around a bright field source which corrupted the image location around the position of FXT\,210423, which persisted after deconvolution. To improve  the quality of our images, we performed iterative self-calibration in both amplitude and phase, which allowed us to recover significantly improved images. We do not detect radio emission from XRT210423 in any of our VLA epochs, but derive constraining upper limits which are summarised in Table \ref{table:VLA}. The corresponding radio luminosity from these observations at four assumed redshift values is shown in Fig. \ref{figure:RadioFXT} along with prompt radio observations from additional FXTs and SGRB radio afterglow data for comparison.

\begin{deluxetable}{cccc}
\tablehead{
\colhead{MJD} \vspace{-0.2cm}  & \colhead{Time since $T_{0}$}& \colhead{Frequency} & \colhead{Flux Density } \\ (d) & \colhead{(d)} & (GHz) & \colhead{($\mu\rm{Jy}\,\rm{beam}^{-1}$)}}
\tablecaption{VLA observations of FXT\,210423. 
Upper limits are 3-$\sigma$ and are measured from a representative region around the source location. 
\label{table:VLA}}
\startdata
59342.0883 & 14.16 & 10 & $<27$\\
59342.1118 & 14.18 & 6 & $<26$\\
59346.3537 & 18.42 & 15 & $<16$\\
59349.0692 & 21.14 & 10 & $<48$\\
59349.0928 & 21.16 & 6 & $<26$\\
59350.2819 & 22.35 & 15 & $<14$\\
59363.2973 & 35.37 & 15 & $<15$\\
59364.0282 & 36.10 & 10 & $<27$\\
59364.0519 & 36.12 & 6 & $<35$\\
\enddata
\end{deluxetable}

In addition to our VLA observations, we include the flux density limits as derived from an observation carried in band-4 (i.e., frequency range 0.55-0.75 GHz) of the Giant Metrewave Radio Telescope (uGMRT) on 2021 June 03 ($\delta t\approx41\,$d under program number DDTC180 (PI Mayuresh Surnis), as reported by \cite{uGMRT_upper_limit}. No continuum emission was detected in the direction of FXT\,210423, and the observation yielded a $3\sigma$ flux density upper limit of $750\mu\rm{Jy}$ at $\nu=0.65$ GHz \citep{uGMRT_upper_limit}.

\subsection{Optical Observations}
\begin{figure}
\includegraphics[width=0.5\textwidth]{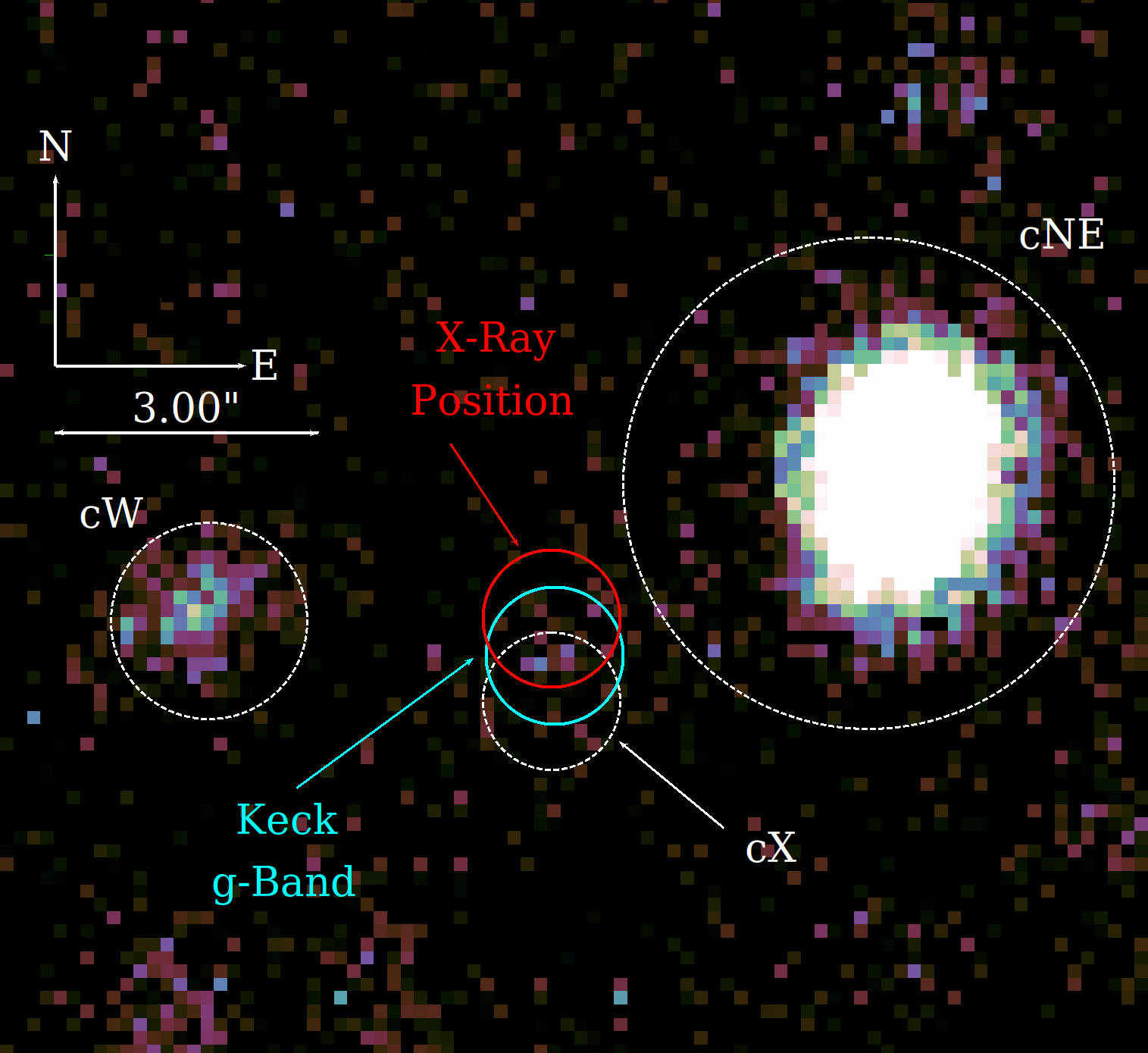}
\caption{Keck $g$-band image of the field of FXT\,210423 acquired on 2021 May 11 ($\delta t=$18.03\,d since discovery). Red region: 0.7$\arcsec$ radius region at the location of the X-ray source from the CXO discovery images as determined with \texttt{wavdetect}. Cyan region: 0.7$\arcsec$ radius region at the location of the extended source that we have identified in our Keck $g$-band images. We also mark the three extended sources (``cX'', ``cW'' and ``cNE'') identified as potential host galaxies by  \cite{eappachen2023fast}. Our Keck source is coincident with ``cX'' and we consider ``cX'' the most likely host galaxy for this FXT.  North is up and East is right. 
}
\label{figure:Optical_hosts}
\end{figure}

We obtained imaging of FXT\,210423 in the $g$ and $i$ filters with the Low Resolution Imaging Spectrograph (LRIS) mounted on the Keck I Telescope on 11 May 2021 which corresponds to $\delta t\approx $18.03 d (PI K.~Paterson, program \#\,O316).  A dithered sequence of 14 frames, each bias and flat-field corrected, were aligned and combined into final deep stacks with total exposure times of 2800 and 2240 seconds in $g-$ and $i-$band, respectively.  A faint source is detected in both filters at a location consistent with the X-ray position of FXT\,210423.  The coordinates of the source in $g$-band are $\rm{RA}=13^{h}48^{m}56.46^{s}$, $\delta=26\degree 39^{'}44.76{''}$ with a $0.3 {''}$ positional uncertainty which corresponds to a $0.36{''}$ offset from the X-ray position of FXT\,210423 and a $0.72{''}$ angular offset from the position of the possible host galaxy proposed by \cite{eappachen2023fast} referred to as ``cX'' hereafter (see Fig. \ref{figure:Optical_hosts}). 
We thus identify our extended $g-$band source with ``cX.'' For ``cX'' \cite{eappachen2023fast} infers $m_{g} = 25.9 \pm 0.1$ mag at $\delta t\approx$ 48 days. We measured the flux of the source in our Keck images using aperture photometry calibrated using SDSS catalog stars, finding AB magnitudes of $m_{g} = 25.80 \pm 0.17$ mag  (consistent with the measurements by \citealt{eappachen2023fast}) and $m_{i} = 25.40 \pm 0.25$ mag. The source is slightly extended in $g-$band making it likely that some of the flux may  be due to an underlying faint host galaxy (with unknown redshift).  In the following, we thus treat these measurements  as upper limits on the optical brightness of a potential optical counterpart of FXT\,210423. 
We use the \cite{fitzpatrick1999correcting} extinction model with $A_v = 0.5422$ mag to correct for Milky-Way reddening.

Additionally, we collected optical observations of FXT\,210423 from the literature \citep{eappachen2023fast, ztfandreoni2021,xinglong2021non, LBTrossi2021non, WaSPandreoni2021p200+}, and we extracted  photometry using the standard Zwicky Transient Facility (ZTF; \cite{Bellm_ZTF, Graham_ZTF} forced photometry service (ZFPS) hosted by IPAC \cite{Masci23} (Appendix \ref{sec:Appendix}).

\section{Constraints on Relativistic Jets} \label{sec:simulations}
\subsection{Jet-afterglow light-curve simulations with BOXFIT}

FXTs have been associated in the literature with manifestations of NS mergers (e.g., \citealt{bauer2017new}), which are known to be capable of launching relativistic jets (e.g., \citealt{Berger14,fong2015decade,ARAAmargutti2021first,Nakar20, Fermi_GRB170817A}). In order to constrain the parameter space of relativistic jets potentially associated with FXT\,210423, we carried out a series of jet-afterglow emission simulations using the publicly available multi-band light-curve generator \texttt{BOXFIT} \citep{BOXFIT}, which is based on two-dimensional hydrodynamics simulations of relativistic jets. 

The jet-afterglow emission detectable after a NS merger is thought to originate from the deceleration of the jet as it interacts with the surrounding medium producing synchrotron radiation \citep{piran2004physicsGRBs}. The evolution of the synchrotron radiation can reveal details about the geometry of the jet (i.e., the jet opening angle $\theta_j$), the observer's angle relative to the jet axis ($\theta_{obs}$), the kinetic energy of the outflow ($E_k$), the density of the surrounding medium ($n$), the fraction of the post-shock energy transferred into magnetic fields ($\epsilon_B$) and relativistic electrons ($\epsilon_e$), and the distribution of energy in the relativistic electrons that produce synchrotron emission (here parameterized as a power-law $N_e( \gamma_e)\propto \gamma_e^{-p}$, where $ \gamma_e$ is the electron Lorentz factor).  Placing constraints on the true energy and the opening angles of these jets can provide insight into the launching mechanism of the relativistic outflow, while constraints on the density of the medium can provide insights on the progenitor model \citep{shibata2019mergerAR, rouco2022jet}. This is especially interesting for FXTs, as their origin is still debated.

We explored a wide range of parameter values (reported in Table \ref{table:Param_summary}) motivated by inferences derived from modeling of SGRB afterglows in the literature (e.g., \citealt{fong2015decade,rouco2022jet}). For all simulations we fixed the value of fractional post-shock energy in relativistic electrons $\epsilon_e = 0.1 $ \citep{sironi2015relativistic}. The power-law index of the electron distribution was fixed to a value consistent with the median value from \cite{fong2015decade} of a sample of 38 SGRBs,  $p=2.4$.  The beaming-corrected jet kinetic energy (i.e. the true energy) $E_{\rm{k}}$ is related to the isotropic equivalent energy as $E_{\rm{k}}=E_{\rm{k,iso}}(1-cos\,\theta_{\rm{j}})$, (e.g., \citealt{frail2001beaming}).
Finally, we ran simulations of the radio, optical and X-ray emission from relativistic jets at four different redshift values to reflect a variety of proposed distances to the source (i.e.,  $z=3.5$, $z=1.51$, $z=1.04$ and $z=0.063$), which we discuss in detail in \S\ref{SubSec:distance}.

\begin{figure*}[!ht]
\includegraphics[width=\textwidth]{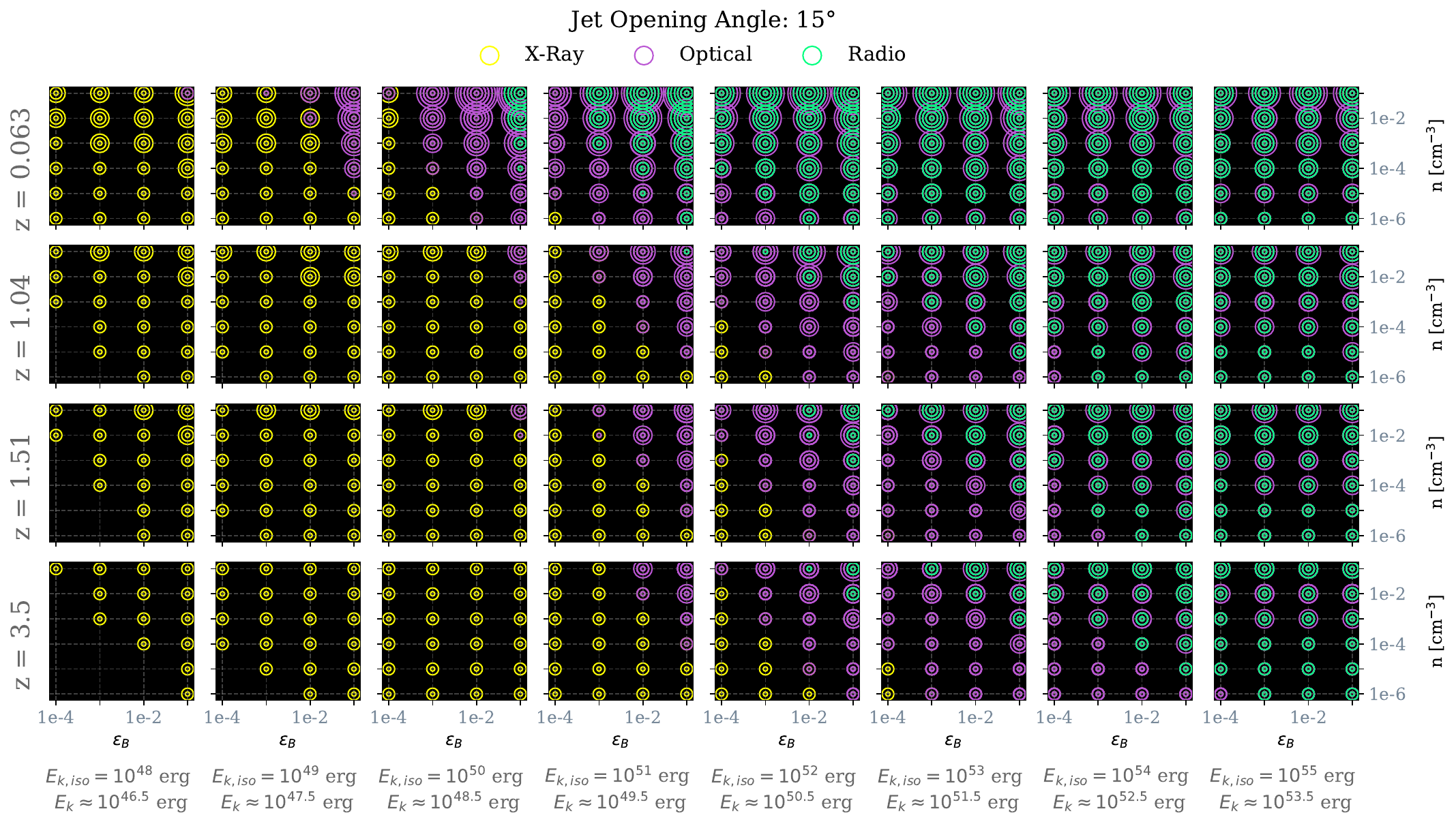}
\caption{Grid representation of simulations that violate observations for a jet with $\theta_j =  15\degree$. Violations occur when either the radio, optical, or X-ray value of an observation exceeds that of the simulation at the corresponding observation time. Colored rings indicate a violation in a particular band (green: radio, purple: optical, yellow: X-ray). The size of the ring indicates  the observation angle of the simulation where the violation occurred with the innermost ring representing a violation of 0$\degree$ and each proceeding concentric ring representing an increase of 10$\degree$. 
A set of 9 concentric rings, in any colour, consequently indicates that any jet with these properties are ruled out. Columns and rows of the the outer grid indicate the energy and distance of the simulation. Columns and rows of each inner box indicate the fraction of the post shock energy transferred into magnetic fields ($\epsilon_B$) and the density of the surrounding medium ($n$) in the simulation. The violation results are overlaid in the order radio (top), optical (middle), and X-ray (last), meaning the appearance of a radio or optical violation may cover an X-ray violation. The ordering has been chosen to best represent the overall shape of the violations. Any crossing on the grid with no ring indicates no constraints could be placed on simulations with the given parameters.  
}
\label{figure:Grid_15_deg}
\end{figure*}

\begin{figure*}[!ht]
\includegraphics[width=\textwidth]{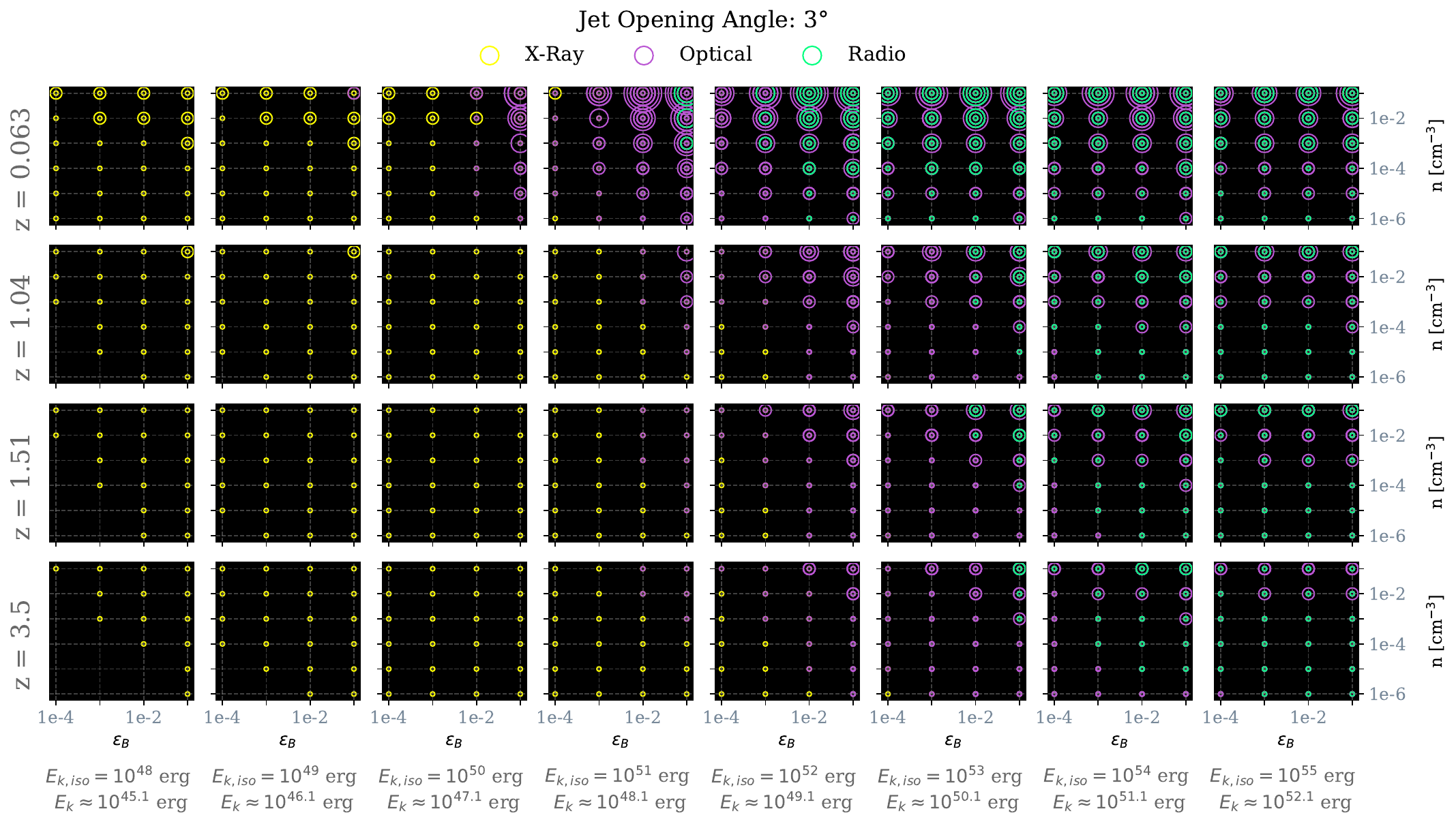}
\caption{Same representation as Fig. \ref{figure:Grid_15_deg} for a jet with $\theta_j=3 \degree$.}
\label{figure:Grid_3_deg}
\end{figure*}

For each combination of the parameters above we simulated the resulting synchrotron emission between  10 - 1000\,d  for observed frequencies $\nu_{\rm{obs}}=0.65, 6, 10, 15$ GHz (radio) and at 10 optical frequencies between $\nu_{\rm{obs}}=(3.143-6.369)\times10^{14}$ Hz. 
For the X-rays we used 1 keV as  the central frequency. For the X-rays and optical we ran simulations starting from the deceleration timescale ($\approx$ few hundred seconds) until 1000\,d.  Each simulated light-curve is compared with our radio and optical  non-detections of the afterglow. We consider the detected X-ray emission as an upper limit on the X-ray brightness of a jet afterglow at that time. A given parameter set is deemed viable if it does not violate any observational constraint. Our results are shown in Fig. \ref{figure:Grid_15_deg} and Fig. \ref{figure:Grid_3_deg} and discussed in \S\ref{sec:disc}.

\begin{deluxetable}{cc}[!ht]
\tablehead{\colhead{Parameter} & \colhead{Values Considered}}
\tablecaption{Adopted values of the parameters for the \texttt{BOXFIT} jet-afterglow simulations. All simulations assume $\epsilon_e=0.1$ and $p=2.4$. \label{table:Param_summary}}
\startdata
$z$ & 0.063, 1.04, 1.51, 3.5 \\
\hline
$\theta_{obs}$ & $0\degree, 10\degree, 20\degree, 30\degree, 40\degree, 50\degree, 60\degree, 70\degree, 80\degree$ \\
\hline
$\theta_{j}$ & $1\degree, 2\degree, 3\degree, 4\degree, 5\degree, 6\degree, 10\degree, 15\degree$ \\
\hline
$E_{\rm{k,iso}}\,\rm{(erg)}$ & $10^{48}, 10^{49}, 10^{50}, 10^{51}, 10^{52}, 10^{53}, 10^{54}, 10^{55}$\\
\hline
$\epsilon_{B}$& $10^{-1}, 10^{-2}, 10^{-3}, 10^{-4}$\\
\hline
$n$  $(\rm{cm^{-3}})$ & $10^{-1}, 10^{-2}, 10^{-3}, 10^{-4}, 10^{-5}, 10^{-6}$\\ 
\enddata
\end{deluxetable}

\pagebreak

\subsection{Distance to the Source} \label{SubSec:distance}

Below we use a number of different arguments to try to infer boundaries on the distance of the FXT\,210423. First, we used our Keck $g$-band detection of an extended source at the location of FXT\,210423 to infer an upper limit on its distance based on the Lyman break.  We found the the limit to be $z < 4.16$, which is notably higher than the typical SGRB redshift of $z\lesssim 1$ \citep{Nugent_SGRB}.
\cite{ai_binary_post_merger_2021} associate the X-ray emission from FXT\,210423 to the emission from a post NS merger magnetar and infer an upper limit on the redshift of the source based on theoretical arguments related to the 
physical details of the origin of the X-ray emission (i.e., ``free zone,'' ``trapped zone,'' or ``jet zone'') and the fraction of magnetar spin-down luminosity dissipated as X-ray emission, $\eta_{x}$. For $\eta_{x} = 10^{-2}$ \cite{ai_binary_post_merger_2021}  derived $z \le 3.5 $, which we adopt here as our high-redshift value.
We adopt two intermediate-redshift values of $z=1.51$ and $z=1.04$ based on the known or photometric redshifts of potential host galaxies. Specifically, three potential host galaxies  ``cX'', ``cW'', and ``cNE'' have been identified (Fig. \ref{figure:Optical_hosts}).  ``cNE'' (i.e.,  SDSS J134856.75+263946.7) has a measured spectroscopic redshift of $z=1.51$ \citep{x-shooterJ, eappachen2023fast}, while ``cW'' is found to have a photometric redshift of $z=1.04$ \citep{eappachen2023fast}. No redshift measurement is available for the most likely host, which is source ``cX'' (coincident with our extended Keck $g$-band source and the X-ray position of the FXT).  

Finally the lowest-redshift value that we consider ($z=0.063$) comes from the initially proposed association of the event with the galaxy cluster Abell 1795   \citep{lin2021chandra} at $d\approx $ 290\,Mpc. While the physical association  with the Abell\,1795 cluster is likely to be due to chance alignment \citep{eappachen2023fast}, we carry out simulations at this redshift to quantitatively demonstrate what constraints can be placed on the presence of relativistic jets in FXTs, were an FXT to be found at these very close distances.

\section{Discussion and Future Prospects} \label{sec:disc} 

We discuss in this section what part of the parameter space of relativistic jets is ruled out by the multi-wavelength observations of FXT\,210423 in the broader context of SGRB-like jets through discussion of  a wide jet ($\theta_j=15\degree$) and a narrow jet ($\theta_j=3\degree$). We end this section with a look at future observational campaigns of new FXTs that are now regularly announced by missions like the Einstein Probe \citep{yuan2022einstein}.  

For our wider jet $\theta_j=15 \degree$ case, Fig. \ref{figure:Grid_15_deg} shows that the entire region of the parameter space is allowed for very off-axis angles $\theta_{obs} > 70 \degree$, which is a consequence of the fainter emission from such jets that is not probed by our observations. Instead, for the roughly on-axis case (i.e., $\theta_{obs} <= 10 \degree$), we can rule out all combinations of parameters aside from the low-energy low-density cases (i.e., $E_{k,iso} = 10^{48}-10^{49}\,\rm{erg}, n=10^{-4} - 10^{-2}\,\rm{cm^{-3}}$) in the lower left of the grid. We note that at intermediate off-axis angles $10 \degree \le \theta_{obs} \le 50 \degree$ we are most sensitive (i.e., our observations rule out a larger portion of the parameter space) for jets with $E_{k,iso} = 10^{50}-10^{52}$\,erg, which is a consequence of the fact that for these jets the time of the observed peak of the emission is better coupled with the time when the observations had been acquired. Finally, we comment on the region of parameter space that is typical of cosmological SGRBs, LGRBs, and TDEs, using the results from \cite{fong2015decade}, \cite{laskar2015energy}, \cite{Alexander20}, and \cite{Eftekhari18}. For typical SGRB energies $E_{k,iso} = 10^{50}-10^{53}$\,erg inferred assuming $\epsilon_B = 0.1$, and typical low-density media with $n=10^{-3}\,\rm{cm^{-3}}$ \citep{fong2015decade}, Fig.\ref{figure:Grid_15_deg} shows that we can rule out all combinations of parameters for jets viewed at a $\theta_{obs} \le 10 \degree$. We conclude that FXT\,210423 did not harbour a wide SGRB-like jet with $\theta_j=15 \degree$ viewed at $\theta_{obs} \le 10 \degree$ even for the largest redshift $z=3.5$ considered in our simulations. 

For the same range of $E_{k,iso}$ values, off-axis, collimated jets will  produce intrinsically fainter emission (because of the lower $E_{k}$), making the detection of such systems more challenging. This is clearly demonstrated by Fig. \ref{figure:Grid_3_deg}, where we present the results for a jet with $\theta_j=3 \degree$. Although we see similar trends to the wider-jet angle case, with more parameter space ruled out at higher energies, higher densities, larger $\epsilon_B$ values and lower distances to the source, for highly-collimated jets with $\theta_j=3 \degree$ we can only rule out on-axis systems for the majority of the parameter space, the exception being the low-energy low-density cases  ($E_{k,iso} \leq10^{49}$\,erg and $n \leq 10^{-2}$ ) in the lower left of the grid. 

LGRBs span a wide range of kinetic energies and densities. We focus our discussion on a representative range of values of  $E_{k,iso} = 10^{50}-10^{52}$\,erg  and $n=10^{-2} - 10^{2}\,\rm{cm^{-3}}$ (inferred for $\epsilon_B = 0.1$, see e.g., \citealt{laskar2015energy}). With reference to Fig. \ref{figure:Grid_15_deg}, we find that  we can rule out all combinations of parameters of jets viewed at a $\theta_{obs} \le 10 \degree$ , bringing us to the conclusion that FXT\,210423 had no LGRB-like jet with $\theta_{j} = 15\degree$ viewed at $\theta_{obs} \le 10 \degree$.

The inferred kinetic energy of prompt on-axis TDE jets from super-massive BHs tend to cover the upper end of the energy range that we have investigated ($E_{k,iso}\ge 10^{52}\,\rm{erg\,s^{-1}}$, e.g., \citealt{Eftekhari18}), with densities $n\ge0.1\,\rm{cm^{-3}}$.\footnote{We acknowledge that our assumed ISM-like density profile might not be representative of density profiles inferred for SMBH TDEs, see e.g., \cite{Alexander20}, their Fig. 2, and that IMBH TDEs might have different energetics than SMBH TDEs.} For this combination of parameters, and assuming equipartition, we can rule out jets viewed at a $\theta_{obs} \le 20 \degree$, bringing us to the conclusion that FXT\,210423 had no TDE jet with $\theta_{j} = 15 \degree$ viewed at $\theta_{obs} \le 20 \degree$. 

We conclude with considerations on future observations of newly discovered FXTs. Our radio observations of the FXT\,210423 demonstrated the constraining power of prompt radio observations of these systems in the first few weeks after discovery. It is clear that rapid radio follow up of future FXTs immediately after discovery can place valuable constraints on the presence and properties of on-axis jets (if there). However, from our results it is also equally clear that  for intrinsically low-energy jets (e.g., highly collimated jets) viewed off axis, very deep, late-time follow up observations on time scales of months to years are necessary.  Sub-$\mu\rm{Jy}$ observations from the Square Kilometre Array (SKA) and next-generation VLA (ngVLA) could be particularly constraining.  The combination of prompt and late-time deep radio observations of the most nearby FXTs discovered by the Einstein Probe will constrain the presence of relativistic jets in these systems and will illuminate their connection (or lack thereof) to NS mergers.

\facilities{VLA, Chandra, Keck}

\software{BOXFIT, GNU}

\section*{Acknowledgments}

The National Radio Astronomy Observatory is a facility of the National Science Foundation operated under cooperative agreement by Associated Universities, Inc. The Keck observations presented in this paper were enabled by access provided by Northwestern University and CIERA. The ZTF forced-photometry service was funded under the Heising-Simons Foundation grant \#12540303 (PI: Graham). This research used the Savio computational cluster resource provided by the Berkeley Research Computing program at the University of California, Berkeley (supported by the UC Berkeley Chancellor, Vice Chancellor for Research, and Chief Information Officer).
R.M. acknowledges support by the National Science
Foundation under award No. AST-2221789 and AST-2224255.
The TReX team at UC Berkeley is partially funded by the
Heising-Simons Foundation under grant 2021-3248 (PI: Margutti). D.L. acknowledges support by the National Aeronautics and Space Administration through Chandra Award No. DD1-22128X issued by the Chandra X-ray Observatory Center and through the ADAP grant No. 80NSSC22K0218. 

\newpage
\appendix
\section{Optical Data}\label{sec:Appendix}

\startlongtable
\begin{deluxetable*}{ccccccc}
\label{appendix:table}
\tablecaption{Optical observations of the field of  FXT\,210423 and inferred brightness at the location of the transient. References: (1) \cite{ztfandreoni2021}; (2) \cite{xinglong2021non}; (3)  \cite{LBTrossi2021non}; (4) \cite{WaSPandreoni2021p200+}; (5) \cite{eappachen2023fast}.}
\tablehead{
\colhead{Facility} & \colhead{Observation} & \colhead{ Days Since $T_0$}& \colhead{Filter} & \colhead{Magnitude}&\colhead{Extinction Corrected} & \colhead{Reference} \\
Observatory & MJD & (days) & & (AB) & Magnitude (AB) &
}
\startdata
Keck & 59345.96 & 18.03 & $g$ & $>25.8$ & $>25.14 $ & This work \\
Keck & 59345.96 & 18.03 & $i$ & $>25.4$ & $>24.97$ & This work \\
ZTF & 59335.37 & 7.44 & $i$ & $>19.6$ & $>19.3$ & [1]\\
ZTF & 59338.34 & 10.41 & $i$ & $>20.4$ & $>20.1$ & This work\\
ZTF & 59341.26 & 13.33 & $i$ & $>20.9$ & $>20.6$ & This work\\
ZTF & 59344.35 & 16.42 & $i$ & $>20.5$ & $>20.2$ & This work\\
ZTF & 59352.37 & 24.44 & $i$ & $>20.3$ & $>20.0$ & This work\\
ZTF & 59353.25 & 25.32 & $i$ & $>20.2$ & $>19.9$ & This work\\
ZTF & 59355.32 & 27.29 & $i$ & $>18.9$ & $>18.6$ & This work\\
ZTF & 59359.34 & 31.41 & $i$ & $>19.8$ & $>19.5$ & This work\\
ZTF & 59362.34 & 34.41& $i$ & $>20.5$ & $>20.2$ & This work\\
ZTF & 59365.37 & 37.44 & $i$ & $>19.9$ & $>19.6$ & This work\\
ZTF & 59328.27 & 0.34 & $r$ & $>20.9$ & $>20.5$ & [1]\\
ZTF & 59329.33 & 1.40 & $r$ & $>20.7$ & $>20.3$ & This work\\
ZTF & 59334.32 & 6.39 & $r$ & $>21.4$ & $>21.0$ & This work\\
ZTF & 59335.29 & 7.36 & $r$ & $>20.8$ & $>20.4$ & This work\\
ZTF & 59336.34 & 8.41 & $r$ & $>21.6$ & $>21.2$ & This work\\
ZTF & 59338.28 & 10.35 & $r$ & $>21.5$ & $>21.1$ & This work\\
ZTF & 59339.27 & 11.34 & $r$ & $>21.3$ & $>20.9$ & This work\\
ZTF & 59340.31 & 12.38 & $r$ & $>21.7$ & $>21.3$ & This work\\
ZTF & 59341.23 & 13.30 & $r$ & $>21.5$ & $>21.1$ & This work\\
ZTF & 59342.35 & 14.42 & $r$ & $>21.6$ & $>21.2$ & This work\\
ZTF & 59343.22 & 15.29 & $r$ & $>21.6$ & $>21.2$ & This work\\
ZTF & 59344.34 & 16.41 & $r$ & $>21.7$ & $>21.3$ & This work\\
ZTF & 59345.26 & 17.33 & $r$ & $>21.4$ & $>21.0$ & This work\\
ZTF & 59346.39 & 18.46 & $r$ & $>21.7$ & $>21.3$ & This work\\
ZTF & 59349.31 & 20.38 & $r$ & $>22.1$ & $>21.7$ & This work\\
ZTF & 59349.28 & 21.35 & $r$ & $>21.6$ & $>21.2$ & This work\\
ZTF & 59350.27 & 22.34 & $r$ & $>21.6$ & $>21.2$ & This work\\
ZTF & 59352.25 & 24.32 & $r$ & $>21.3$ & $>20.9$ & This work\\
ZTF & 59353.30 & 25.37 & $r$ & $>21.0$ & $>20.6$ & This work\\
ZTF & 59354.24 & 26.31 & $r$ & $>21.1$ & $>20.7$ & This work\\
ZTF & 59355.23 & 27.30 & $r$ & $>20.3$ & $>19.9$ & This work\\
ZTF & 59356.28 & 28.35 & $r$ & $>20.8$ & $>20.4$ & This work\\
ZTF & 59358.32 & 30.39 & $r$ & $>20.3$ & $>19.9$ & This work\\
ZTF & 59359.22 & 31.29 & $r$ & $>20.6$ & $>20.2$ & This work\\
ZTF & 59361.35 & 33.42 & $r$ & $>20.8$ & $>20.4$ & This work\\
ZTF & 59362.25 & 34.32 & $r$ & $>21.3$ & $>20.9$ & This work\\
ZTF & 59363.34 & 35.41 & $r$ & $>21.2$ & $>20.8$ & This work\\
ZTF & 59364.30 & 36.37 & $r$ & $>21.6$ & $>21.2$ & This work\\
ZTF & 59328.34 & 0.41 & $g$ & $> 20.5 $ & $>19.8$ & [1]\\ 
ZTF & 59329.27 & 1.34 & $g$ & $> 20.1 $ & $>19.4$ & This work\\
ZTF & 59331.47 & 3.54 & $g$ & $> 19.4 $ & $>18.7$ & This work\\ 
ZTF & 59334.22 & 6.29 & $g$ & $> 22.0 $ & $>21.3$ & This work\\
ZTF & 59335.29 & 7.36 & $g$ & $> 21.3 $ & $>20.6$ & This work\\
ZTF & 59336.34 & 8.41 & $g$ & $> 21.3 $ & $>20.6$ & This work\\
ZTF & 59338.36 & 10.43 & $g$ & $> 21.8 $ & $>21.1$ & This work\\
ZTF & 59339.23 & 11.3 & $g$ & $> 21.6 $ & $>20.9$ & This work\\
ZTF & 59340.36 & 12.43 & $g$ & $> 21.6 $ & $>20.9$ & This work\\
ZTF & 59341.3 & 13.37 & $g$ & $> 21.8 $ & $>21.1$ & This work\\
ZTF & 59342.34 & 14.41 & $g$ & $> 21.2 $ & $>20.5$ & This work\\
ZTF & 59343.27 & 15.34 & $g$ & $> 21.7 $ & $>21.0$ & This work\\
ZTF & 59344.31 & 16.38 & $g$ & $> 22.2 $ & $>21.5$ & This work\\ 
ZTF & 59345.20 & 17.27 & $g$ & $> 21.8 $ & $>21.1$ & This work\\
ZTF & 59346.35 & 18.42 & $g$ & $> 21.8 $ & $>21.1$ & This work\\
ZTF & 59348.34 & 20.41 & $g$ & $> 22.2 $ & $>21.5$ & This work\\
ZTF & 59352.20 & 24.27 & $g$ & $> 21.0 $ & $>20.3$ & This work\\
ZTF & 59353.34 & 25.41 & $g$ & $> 21.1 $ & $>20.4$ & This work\\
ZTF & 59354.33 & 26.40 & $g$ & $> 21.0 $ & $>20.3$ & This work\\
ZTF & 59355.30 & 27.37 & $g$ & $> 20.3 $ & $>19.6$ & This work\\
ZTF & 59356.22 & 28.29 & $g$ & $> 20.7 $ & $>20.0$ & This work\\
ZTF & 59358.40 & 30.47 & $g$ & $> 20.1 $ & $>19.4$ & This work\\
ZTF & 59359.27 & 31.34 & $g$ & $> 20.3 $ & $>19.6$ & This work\\
ZTF & 59361.29 & 33.36 & $g$ & $> 20.8 $ & $>20.1$ & This work\\
ZTF & 59362.21 & 34.28 & $g$ & $> 21.0 $ & $>20.3$ & This work\\
ZTF & 59363.26 & 35.33 & $g$ & $> 21.3 $ & $>20.6$ & This work\\
ZTF & 59364.29 & 36.36 & $g$ & $> 21.1 $ & $>20.4$ & This work\\
ZTF & 59365.29 & 37.36 & $g$ & $> 21.1 $ & $>20.4$ & This work\\ 
LBT & 59342.31 & 14.38 & $z$ & $>25.1$ &$>24.88$& [3]\\
LBT & 59342.31 & 14.38 & $r$ & $>26.1$ & $>25.66$ & [3]\\
Xinglong & 59339.53 & 11.60 & $I$ &  $>20.5$ & $>20.2$ & [2]\\
Palomar WaSP & 59340.30 & 12.37 & $i$ & $>24.8$ & $>24.32$ & [4]\\
Palomar WaSP & 59340.30 & 12.37 & $r$ & $>25.2$ & $>24.76$ & [4]\\
VLT & 59340 & 13 & $R$ & $>24.7$ &$>24.29$ & [5]\\
GTC &  59375 & 47 & $u$ & $>26.2$ & $>25.35$ & [5]\\
GTC &  59375 & 47 & $g$ & $>27.0$ & $>26.56$ & [5]\\
GTC&  59375 & 47 & $r$ & $>26.1$ & $>25.66$ & [5]\\
GTC &  59375 & 47 & $i$ & $>24.4$ & $>24.08$ & [5]\\
GTC &  59375 & 47 & $z$ & $>24.7$ & $>24.46$ & [5]\\
\enddata
\end{deluxetable*}

\bibliography{BibliographyFile}{}
\bibliographystyle{aasjournal}

\end{document}